\documentclass[useAMS,usegraphicx]{mn2e}
\input epsf.tex

\title[What goes around, comes around] {The circulation of dust in protoplanetary discs and the initial conditions of planet formation}
\author[Bradley M. S. Hansen]{Bradley M. S. Hansen$^{1}$\thanks{E-mail:hansen@astro.ucla.edu}\\
$^{1}$Department of Physics \& Astronomy, University of California Los Angeles, Los Angeles, CA 90095}
\begin{document}

\date{submitted Dec 2013}

\pagerange{\pageref{firstpage}--\pageref{lastpage}} \pubyear{2014}

\maketitle

\label{firstpage}

\begin{abstract}

We examine the consequences of a model for the circulation of solids in a protoplanetary nebula in which
aerodynamic drag is counterbalanced by the recycling of material to the outer disc by a protostellar outflow or a disc wind.
This population of circulating dust eventually becomes unstable to the formation of planetesimals by gravitational
instability, and results in the ultimate deposition of $\sim$ 30--50 $M_{\oplus}$ in planetesimals on scales $R< 1 AU$. Such
a model may provide an appropriate justification for the approximately power law initial conditions needed to reproduce observed
planetary systems by in situ assembly. 

\end{abstract}


\section{Introduction}

The detection of planets orbiting other stars has revealed a great diversity of both planetary mass and location,
including several populations which have no analogue in our own solar system. Of particular interest is the
discovery of substantial numbers of sub-Jovian planets with orbital periods shorter than that of Mercury, 
using both the radial velocity and transit techniques (Howard et al. 2010; Mayor et al. 2011; Borucki et al. 2011; Batalha et al. 2013). 

This population proved to be a surprise for models of planetary systems whose short period populations are
generated by migration inwards from larger radii (e.g. Ida \& Lin 2008). However, the properties of the observed
systems do match the expectations of simple in situ assembly models (Hansen \& Murray 2012, 2013; Chiang \& Laughlin 2013), although
the amount of mass in rocky material required for such models is sometimes larger than what one might anticipate from a simple
minimum-mass solar nebula model. Hansen \& Murray (2012) suggested that such conditions could be realised if solid material is concentrated radially
in the inner parts of the gas disc prior to the late-stage assembly into solid planets. This is not an outrageous expectation as it is
well known that small bodies in gas discs are potentially subject to dynamically important
aerodynamic drag forces (Whipple 1972; Weidenschilling 1977a; Takeuchi \& Lin 2002; Bai \& Stone 2010). However, the particular details of how
such a model might set the stage for planet formation are still unclear.
 In this paper we attempt to outline a simple model that provides such a framework.

\section{Trapping versus Circulation}

Several authors have discussed the possibility that planet formation may be seeded by special locations in the
disc which allow radially migrating material to collect and assemble (Haghighipour \& Boss 2003; Masset et al. 2006;
Hasegawa \& Pudritz 2011; Kretke \& Lin 2012). 
Recently, two groups of authors (Boley \& Ford 2013; Chatterjee \& Tan 2013) have proposed models that use
this idea as the basis for a scenario for inside-out 
planet formation, in which solid material migrates inward due to drag forces until it 
 it stalls due to a pressure maximum, either at the inner edge of the dead zone or near an evaporation
or sublimation front. The resulting accumulation of
material causes planetesimals to form and to extend the truncation of the disc outwards, resulting in what
these authors call `Systems with Tightly-packed Inner Planets', or `STIPS'. These proposals may indeed provide
compact planetary systems, but they may, in fact, produce systems that are too radially concentrated.

\begin{figure}
\includegraphics[width=84mm]{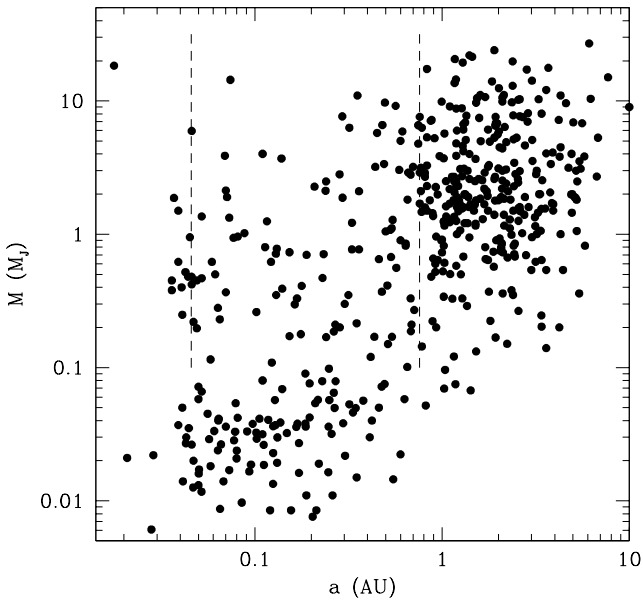}
\caption[am.ps]{The points show the various planetary masses (modulo an unknown
inclination angle) detected by radial velocities only (transit detection exhibits a strong
selection function that may influence this discussion), as of October 2013. The vertical dashed lines indicate
two locations where the incidence of giant planets appears to change dramatically,
a possible indication of the physics that determines the planet formation and
evolution. No such obvious change in planet frequency appears for masses $<0.1 M_J$.
\label{am}}
\end{figure}

 The distribution of known extrasolar planets (Figure~\ref{am}) shows 
two distinct length scales for giant planets (an overdensity near 0.05~AU and an increase in frequency $>0.7$~AU) but
no corresponding special location for planets with mass $< 0.1 M_J$, as might be expected if the pile-up
is generated by a feature in the underlying disc structure, such the inner edge of a dead zone. Indeed,
attempts to characterise these planetary systems consistently find relatively smooth distributions.
Chiang \& Laughlin (2013) construct a 
`Minimum Mass Kepler Nebula' by considering the ensemble properties of the Kepler planet candidate sample, and
derive an empirical distribution consistent with an  power-law distribution of surface density, $\Sigma \propto R^{-1.6}$.
Hansen \& Murray (2012, 2013) recover a very similar underlying model in matching their assembly simulations for these planets
to the observations, and found that initial density profiles that were too steep produced planetary systems that were
too compact to match the observations.

The inside-out packing of planets in the STIPS models produce a spacing of planets that is dictated by the considerations
of dynamical stability of interacting neighbouring planets (Chatterjee \& Tan 2013). Monte Carlo models of the in situ
assembly of the observed planetary systems (Hansen \& Murray 2013)  
 are able to match the observed distributions of periods
and period ratios of the observed Kepler systems. However, the resulting distribution in planetary spacings peaks at
factors of 2--3 times larger than the threshold for dynamical stability, suggesting that the final configuration is
not as tightly packed as it would be if it were assembled sequentially from the inside out.

In this paper, we consider an alternative notion -- namely that the solid material is not trapped, but rather 
migrates rapidly to the inner edge of the protoplanetary disc, from whence it is
 returned to the outer disc, to begin the inspiral anew. 
We then examine the consequences of this
hypothesis and the manner in which the resulting steady-state distribution of solids can form the 
initial conditions for the later stages of planetary formation. 
Such a model finds support within our own Solar System via  the
evidence for widespread processing at high temperatures in  chondrites (e.g. Scott \& Krot 2005) 
and 
 in the observation that the chemical constituents
of the comet Wild-2 show evidence of substantial chemical processing in the hotter inner regions of the protosolar
nebula (Brownlee et al. 2006; McKeegan et al. 2006; Brownlee, Joswiak \& Matrajt 2012).

Indeed, such considerations have long spurred suggestions that material was recirculated in the protosolar 
nebula (Liffman \& Brown 1995; Shu, Shang \& Lee 1996; Shu et al. 2001; Hu 2010; Salmeron \& Ireland 2012).
These models   
propose that solid material in the inner parts of the protoplanetary nebula are levitated and carried outwards
by either the magnetic funnel of the stellar wind or by a magnetically driven outflow from the disk, ultimately
 to be returned to the protoplanetary nebula on larger scales.
Protostellar outflows are a common feature of young stars, and the entrainment of solid material
into such `winds' offers a mechanism to redistribute material from the locations near the
star to more distant locations. The most prominent of these models is that of Shu et al., which proposes that the chemical processing
observed in the constituents of asteroids (at $\sim 3$~AU from the Sun) may have, in large part,
occurred at distances of a few stellar radii, with the material subsequently swept up into the
outflow until the material decoupled from the wind and fell back to the disc.
 One of the
principal motivations of this model was to describe the heating and chemical alteration of metereotic 
components such as chondrules by processes near the magnetic X-point where the disc and the stellar magnetosphere
meet. Recent cosmochemical and age-dating evidence (Connelly et al. 2008; Desch et al. 2012) favor more traditional
explanations for these effects, but this does not necessarily invalidate the dynamical component of the X-wind model.
An alternative launching mechanism can be found in models where the inner parts of protoplanetary discs are
dominated by strong magnetic 
fields which launch magnetocentrifugal winds (K\"{o}nigl \& Pudritz 2000; Bai \& Stone 2013; Armitage, Simon \& Martin 2013)
from a wider range of radii, 
which may also entrain solid materials and return them to the outer disc. 

Our goal is to examine the effect that such a model has on the global inventory of solid material in the protoplanetary
disc, and how it might set the stage for subsequent evolution.
Thus, in \S~\ref{Model}, we describe a simple model of an evolving gaseous protoplanetary disc, within which we
describe the dynamics of small (dust) particles of various size, which migrate radially inwards until they reach
the inner edge of the disc, at which point they are redistributed outwards to begin the cycle again. In \S~\ref{Results}
we describe the resulting evolution of gas and dust components and how they set a context for planet formation through
the onset of gravitational instability and the formation of planetesimals.

\section{Gas Disk Model}
\label{Model}

In order to describe the dynamics of the solid particles, we need to adopt a model for the evolution of the
gaseous protoplanetary disc, which acts as the background for the evolution of the solids. We follow the description of Alexander \& Armitage (2007)--hereafter AA07-- which provides a 
simple model of a protoplanetary disc that includes the viscous evolution as well as the evolution through
the transition disc phase and eventual dissipation of the disc.

The gas surface density $\Sigma_g$ evolves according to the traditional one-dimensional diffusion equation
\begin{eqnarray}
\frac{\partial \Sigma_g}{\partial t} = \frac{3}{R}\frac{\partial }{\partial R} \left[
R^{1/2} \frac{\partial}{\partial R} \left( \nu \Sigma_g R^{1/2} \right) \right] - \dot{\Sigma}_{w},
\end{eqnarray}
where $R$ is the cylindrical radius, $\nu$ is the kinematic viscosity and $\dot{\Sigma}_{w}$ is the mass
loss rate due to a wind from the disc, which is a function of $R$ and time $t$. The viscosity is based on the $\alpha$ prescription
$\nu (R) = \alpha \Omega H^2$, where $\Omega$ is the orbital frequency and $H(R)$ is the disc scaleheight.
Following AA07, we adopt
$ H(R) \sim 0.0333 (R/1 AU)^{5/4}$, which yields a viscosity that grows linearly with $R$. We will assume
$\alpha=0.01$ as our default viscosity parameterisation.

The AA07 model also includes a disc wind, driven by photoionization from the central star. The inclusion
of the evaporation yields an evolution that begins with the standard accretion and viscous spreading,
but eventually results in a gap opened in the gas disc at $R \sim 1$AU. At this point the gas interior  
to the gap accretes rapidly onto the star, leaving an outer disc and an inner hole. The outer disc is
then slowly removed by evaporation. For simplicity, we 
  restrict our model to the diffuse radiation field used in AA07, as 
the additional direct radiation component applies to the later dissipation of the outer
disc once the inner disc has been evacuated, a phase of evolution of less interest to us. We use the 
wind profile as outlined in the appendix~A of AA07.

An alternative mode of disc clearing is presented by Armitage, Simon \& Martin (2013) -- ASM hereafter -- motivated by studies
which suggest that the conservation of magnetic flux in a disc leads to an increase in the strength of the
magnetic disc wind as the gas surface density decreases. This in turn drives an evolution of the viscosity
parameter $\alpha$, and an increase in the rate of viscous evolution with time. This can result in a 
qualitatively different clearing, by removing the outer disc first and then finally the inner disc. We incorporate
this as an alternative disc model by neglecting the photoevaporation but including an evolution of $\alpha$ as
parameterised by ASM. This evolution is based on the global conservation of magnetic flux in the
disc, which amounts to the conservation of the integral of $r^{1/8} \Sigma^{1/2}$ in our model.

Our default initial disc is 
\begin{equation}
\Sigma_g = \frac{1045 g/cm^2}{R/1 AU} \left( 1 - \sqrt{ \frac{0.01 AU}{R}} \right) exp \left( - \frac{R}{20 AU} \right) \label{Init}
\end{equation}
which is of similar form to that of AA07 but more compact, as befits our interest in the inner regions of the disc.
 Our model for the initial deposition of solid particles
is drawn from Hughes \& Armitage (2012), setting the condensible mass fraction to be $0.0049$ interior to the
water ice line (which we take to be at 3~AU), 0.0106 between the water ice line and the methane ice line
(taken to be at 15 AU), and 0.0149 exterior to that. We initially neglect grain growth and so we assume all the condensible
material is deposited at the beginning and we thereafter simply follow the dynamical evolution. We will discuss growth
effects in \S~\ref{Sizes}.

\section{Solid Particle Evolution}

The solid particles that condense out of the gas phase must eventually grow 
into larger bodies to form planets. However, there are several significant hurdles along the path to planet
formation. The first is that, as small particles grow larger, their coupling to the gas weakens and
the resulting difference in orbital velocity means that they are subject to aerodynamic drag forces.
This causes them to spiral
inwards towards the star (Whipple 1972; Weidenschilling 1977a), where they may be lost. We incorporate this into our model 
 using the formalism of AA07, based
largely on Takeuchi \& Lin (2002),
in which the radial velocity of each particle is given by
\begin{equation}
u_d = \frac{\tau^{-1} u_g - \eta u_k}{\tau + \tau^{-1}}.
\end{equation}
The quantity $u_g$ represents the radial velocity of the gas due to the viscous evolution, and $u_k$ is the Keplerian orbital
velocity at distance $R$. Thus, as the parameter $\eta$ (defined below) becomes non-negligible, the solids
will begin to evolve radially with respect to the gas.
 The non-dimensional stopping time $\tau$ is given by
\begin{equation}
\tau = \frac{\pi \rho_d s}{2 \Sigma_g} \label{EpsTau}
\end{equation}
over most of the disc,
where $\rho_d$ and $s$ are the density and radius of a given grain. This quantity determines whether the 
particles are well coupled to the gas ($\tau<1$) or not. The quantity $\eta$ reflects the differential motion
between a particle in a true Keplerian orbit and the gas, which is partially supported by a radial pressure
gradient,
\begin{equation}
\eta = - \left( \frac{H}{R} \right) \left( \frac{\partial \log \Sigma_g}{\partial \log R } - \frac{7}{4} \right).
\end{equation}
The nature of the drag force changes depending on the size of the particle relative to the mean free path in
the gas. Over the bulk of our disc, the drag force is of the Epstein form (equation~\ref{EpsTau}). However, in
the outer part of the disc, at late times, it can 
become of Stokes form
\begin{equation}
\tau = \frac{8 m_p}{9 \sigma_{H2} \Omega} \frac{\rho_d s^2}{ \Sigma^2_g}, 
\end{equation}
where $m_p$ is proton mass, $\sigma_{H2}$ is the collision cross-section of molecular Hydrogen\footnote{Taken to be $2 \times 10^{-15} cm^2$ -- see discussion
in Chiang \& Youdin (2010)}, and $\Omega$ is the
orbital frequency.

We must also account for diffusive effects due to gas turbulence, which acts against particle density gradients
established by differential drag. Thus, we include a diffusive velocity term 
\begin{equation}
u_{diff} = - D \frac{\partial}{\partial R} \ln \left[ \frac{\Sigma_{d}}{\Sigma_g} \right]
\end{equation}
where $\Sigma_d$ is the local dust density, and $D$ is the turbulent diffusivity. The default assumption is to
equate this with the viscosity $\nu$ (defined as the Schmidt number $\nu/D = 1$). Deviations from unity
are possible, but our focus is on the smaller particles, while proposed corrections are
most important for larger particles in the marginally coupled limit, as discussed by Youdin \& Lithwick (2007). We assume the diffusion coefficient is the same for all particle sizes, but that the diffusion
velocity of each particle size is determined by the surface density of that species only.

\subsection{Redistribution}

Our model thus far represents a standard treatment of the evolution of both gas and solids in an evolving
protoplanetary disc. To this,
 we now add 
 the hypothesis that solid material is entrained by a wind or jet in the
inner disc (Liffman \& Brown 1995; Shu et al. 1996; 2001; Salmeron \& Ireland 2012; Bai \& Stone 2013) and transferred back to the outer disc. 
The original Shu et al. model focussed primarily on redistrbution only as far as the asteroid
belt, but 
Hu (2010), motivated by the StarDust results, reviewed the aerodynamic requirements for  redistribution in an X-wind model and found
that, with uncertainties due to outflow and disc geometries, particles of size in the range of
microns to millimeters could be transferred as far as the Kuiper Belt regions of discs. Larger particles are
less well coupled to the outflow and likely to return to the disc on smaller scales, while smaller
particles are sufficiently well entrained that they are easily lost from the system entirely. Thus,
it is possible to invoke such models to explain the observed composition of the Comet Wild-2 as well.

We incorporate these effects into our simple model by assuming that,
 when solid particles reach 0.05~AU, they are instantly
transferred back to the outer disc, with an equal probability per unit area of landing between 35 and 90~AU. This keeps
 the entire solid inventory circulating between the inner and outer disc, although the gas disc is slowly
being accreted onto the star or evaporated. We will examine the effects of varying the properties of the
redistribution in \S~\ref{Loss}.

\subsection{From Dust to Planetesimals}

Our hypothesis of wind entrainment and circulation of solid material provides a potential solution to the first
hurdle to planet formation, in that it prevents the accretion of the solid material by the star.
There is a second well-known hurdle to planet formation, which is the difficulty of converting small, dust-size particles into larger bodies
with the densities and tensile properties of planetesimals. Despite decades of work, there is little convincing
evidence that this transformation can be achieved by incremental processes, and thus the best hope for forming
planetesimals is the gravitational instability of a solid dust layer that sediments to the midplane (Goldreich \& Ward 1973).
However, this model has also proven difficult to implement because turbulent gas motions can prevent a sufficiently dense
dust layer from forming (Weidenschilling 1980). In order for the solid surface
densities to be high enough to overcome turbulent stirring, Youdin \& Shu (2002) demonstrated that the local
solid densities had to be enhanced by factors of at least several over traditional protoplanetary metallicities.

There have been 
  a variety of proposals (see Chiang \& Youdin 2010 for a review) for achieving such
enhancements. One class of models rely on the backreaction on the gas from radial particle streaming (Youdin \& Chiang 2004; Youdin \& Goodman 2005; Johansen et al. 2007) to
promote gravitational instability. However, this only works if the particles are big enough to decouple from the gas on timescales similar to the orbital
time ($\tau \sim 1$), which requires growth to sizes well in excess of the micron--centimeter sizes considered here. Another set of proposals includes growth of
secular modes (Ward 2000; Youdin 2011) to produce particle clumping. However, these models also require either large particles or low turbulent viscosities ($\alpha < 10^{-4}$)
to produce realistic growth.

 In our simple model, the global dust inventory remains in
circulation, even while the disc slowly loses gas by accretion and wind loss. As a result, the traditional scenario for
direct gravitational instability remains potentially viable, because the
 required overdensities will eventually come to pass, as the global solid/gas ratio is monotonically increasing, 
although the locations where the instability occurs will depend on the evolutionary details. These locations are what
is of interest to us, as they will then dictate the initial conditions for the later stages of planetesimal accumulation.

Therefore, we adopt the Youdin \& Shu (2002) -- YS02 -- model of planetesimal formation. Interpreting the YS02 critical dust
surface density in terms of our disc model yields an expression for the critical dust surface density 
\begin{equation}
\Sigma_d = 0.033 g.cm^{-2} \Sigma_g \left( \frac{R}{1 AU} \right)^{1/4} s(\psi)
\end{equation}
where $s(\psi) = (1+\psi) \ln \left[ (1+\psi+\sqrt{1+2 \psi})/\psi \right] - \sqrt{1 + 2 \psi}$, and
$\psi = 4.3 \times 10^{-5} \Sigma_g (R/AU)^{7/4}$. As an example, for a gas surface density of 
$\Sigma_g = 100 g.cm^{-2}$ at $R= 1AU$, this requires $\Sigma_d/\Sigma_g=0.18$ for gravitational
instability. This represents  an enhancement factor of 37 over the original, in situ, ratio, although the precise value of the
necessary enhancement may be uncertain (e.g. Gomez \& Ostriker 2005; Shi \& Chiang 2013).

Whenever the local surface density of dust exceeds the critical value, we assume the excess is transformed
into planetesimals by gravitational instability and no longer evolves with drift or with the gas. This
assumption is a good one until the planetesimals assemble into larger bodies, at which point they may
migrate again by exchanging angular momentum with the disc gas. We shall assume that this latter process
takes $> 5$Myr -- the point at which our calculation ends. Therefore, the final result of our calculation is a deposition
profile of planetesimals that result from the global evolution of the dust and gas populations, which may then
serve as the initial conditions for the later assembly of larger bodies.

\section{Results}
\label{Results}

\subsection{No Recycling}
\begin{figure}
\includegraphics[width=84mm]{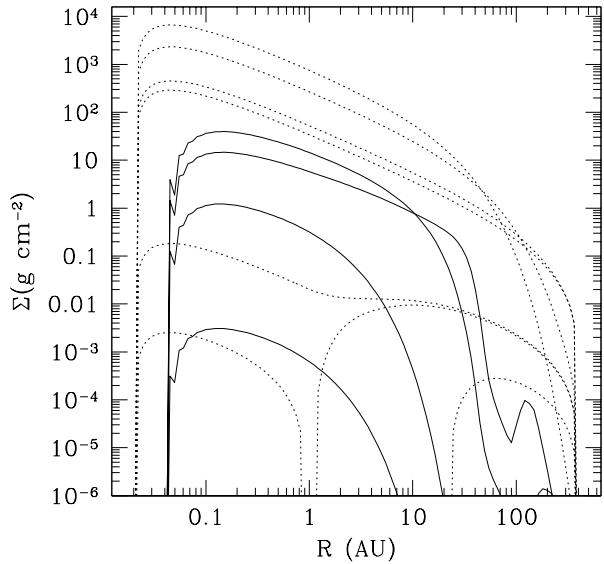}
\caption[Ref.ps]{The top four dotted curves indicate the gas surface density profile at ages
of 0, 0.1, 0.5 and 0.7~Myr respectively. The four solid curves indicate the surface density of
a population of 1~mm dust grains at the same ages. We see that the dust is completely removed
within 1~Myr in the absence of other effects. The bottom three dotted curves indicate the late
time evolution of the gas disc, at ages 4.0, 4.04 and 5~Myr, and show the rapid depletion of the
inner gas disc once photoevaporation becomes important.
\label{Ref1}}
\end{figure}

We begin with a reference model, 
 in which the gas disc given by equation~(\ref{Init})
is evolved, along with our model for dust drift, diffusion and formation of planetesimals included. However,
in this reference model, there is no redistribution of material from the inner disc to the outer disc. Thus,
the reference model represents the expectations for planetesimal formation if we adopt the standard
picture of viscous gas evolution and dust drift. This model also requires a choice of the size of solid particles, as they will determine
the overall drift rates. 
Figure~\ref{Ref1} shows the evolution of a dust population of size 1mm, where particles are 
simply removed when they reach the inner edge. The dust population is substantially depleted
within 0.3~Myr and completely removed within 1~Myr. Other sizes behave in a similar fashion,
albeit with different timescales -- larger dust is removed more rapidly and
smaller dust is retained longer.

The gas disc shows the expected evolution, with the bulk of the mass moving inwards and accreting onto the
star, with a small fraction moving outwards to conserve angular momentum. We see that the effect of the
wind starts to have an influence after $\sim 4$~Myr, and that this rapidly causes a gap to form at
$R \sim 1 AU$, and the gas interior to this is quickly accreted onto the star. The outer gas disc is
slowly eaten away from the inside out by the wind mass loss. 

The radial drift of the dust will enhance solid surface densities and potentially generate
sufficient overdensities for planetesimal formation (Youdin \& Chiang 2004), but studies in
evolving gas discs have not found significant mass deposition (Hughes \& Armitage 2010). In
our model too, we find little planetesimal formation. Our models do satisfy the analytic
criteria for particle pile-up identified by Youdin \& Chiang, but the enhancements are not
sufficient to generate gravitational instability without recycling (however, see \S~\ref{Wind}).

\subsection{Simple Circulation}
\label{SingleSize}

\begin{figure}
\includegraphics[width=84mm]{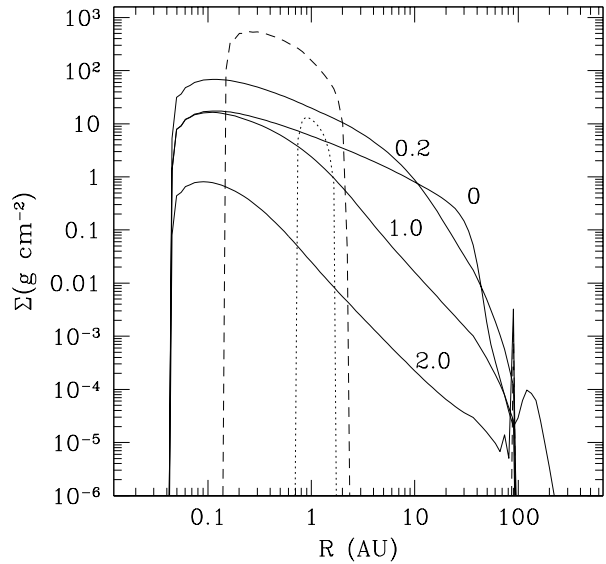}
\caption[X1.ps]{The solid curves indicate the surface density profile of the
dust at ages of 0, 0.2, 1 and 2 Myr. All dust in this model has the size 1mm and
evolves in the same gaseous background as shown by the dotted lines in Figure~\ref{Ref1}.
The dotted line indicates the surface density of planetesimals at 0.2~Myr (early deposition)
and the dashed line the same but at 4~Myr (final deposition).
\label{X1}}
\end{figure}

The essential problem is that the radial migration of dust is too rapid in a
protoplanetary disc for easy planetesimal formation. Figure~\ref{X1} now demonstrates
the effect of including our model for redistribution of the solid material back to
the outer disc. This model uses the same parameters as the model shown in Figure~\ref{Ref1},
with dust of size 1~mm.
 After several radial travel times, the dust establishes a
steady-state surface density profile in the disc, which is steeper than
the original dust profile. As the gas disc evolves and the gas surface
density decreases, the dust profile will eventually exceed the critical density at particular
locations and thus planetesimals will form through gravitational instability.
 The
deposition of planetesimals begins at $\sim 1$~AU but eventually extends inwards to
$\sim 0.2$~AU. The steep dust density profile can be understood by the
dynamics of dust in the $\tau \ll 1$ limit, which leads to a drift velocity
which scales $\propto R^{3/4}$ in the case where the gas surface density $\propto 1/R$.
Given this density profile, the dust population achieves steady state flux when
$\Sigma_d \sim 1/{R u_d} \sim R^{-7/4}$.

\begin{figure}
\includegraphics[width=84mm]{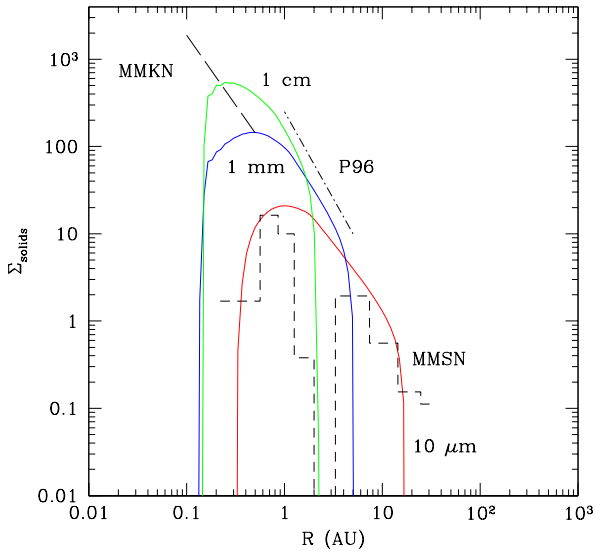}
\caption[Solid2.ps]{The dashed line labelled MMKN is the surface density
profile advocated by Chiang \& Laughlin (2013) as a Minimum Mass Kepler Nebula.
The dashed histogram labelled MMSN is the minimum-mass solar nebula of Weidenschilling
(1977b), although we have restricted ourselves to an inventory of the solid material
in the Solar System (assume 5 $M_{\oplus}$ cores for the outer Solar System planets).
The dot-dashed line is the solid surface density estimate by Pollack et al. (1996) for
the formation of giant planet cores.
The green curve represents the planetesimal deposition in our model for a dust
component composed of 1~cm (large) particles, the blue curve is for a model in
which particles have size 1~mm, and the red curve shows the equivalent
result when the particles are composed of 10$\mu$m (small) particles, which move through
the disc more slowly. 
\label{Solid2}}
\end{figure}

The speed with which material moves radially through the disc is the result of
particle size, and so the normalisation of the steady state surface density profile varies with the
size of the particle (the overall slope is the same as long as $\tau \ll 1$).
 Figure~\ref{Solid2} shows the final planetesimal deposition
profiles  of models in which all the
particles have radii of $s=1cm$, $s=1 mm$ and $s=10 \mu m$ respectively. Also
shown are estimates of the required rocky inventories for the planetary systems
seen by radial velocity (RV) studies and Kepler, as well as for our own terrestrial and giant planets.
For the inner disc we use the estimate of Chiang \& Laughlin (2013), while we
use a modification of the Weidenschilling (1977b) estimate of the minimum mass
solar nebula (in that we restrict ourselves to the rocky component here).
We see that circulating populations of small particles in the $\tau \ll 1$ regime
are able to provide an inventory of planetesimals that matches the requirements imposed
by both terrestrial and extrasolar observations. Note in particular that this model
provides a rationale for the smooth, quasi-power-law distributions usually adopted
in planet formation models.

\subsection{Size Distribution}
\label{Sizes}

A proper model for the evolution of the small bodies would also include 
a model for particle growth over time, which would couple the chemistry and
the dynamics of small bodies in the disc. Such a calculation is well beyond
our simple model, but we can anticipate some of the consequences of grain
growth by considering
a composite model of sizes designed to examine the differential motion of both small
and large particle components in \S~\ref{SingleSize}. 

Our single size distribution shown in Figure~\ref{X1} was chosen on the basis
that the evidence from solar system meteorites suggests that the size distribution
of chondrules peaks between 0.1--1~mm (Scott \& Krot 2005) and thus a dust 
population of such particles offers a reasonable simple model of the underlying
constituents. However, if this is the result of aerodynamic sorting in the
formation of the chondrites (e.g. Cuzzi, Hogan \& Shariff 2008), then the original
size distribution of the nebular solids may be different.

\begin{figure}
\includegraphics[width=84mm]{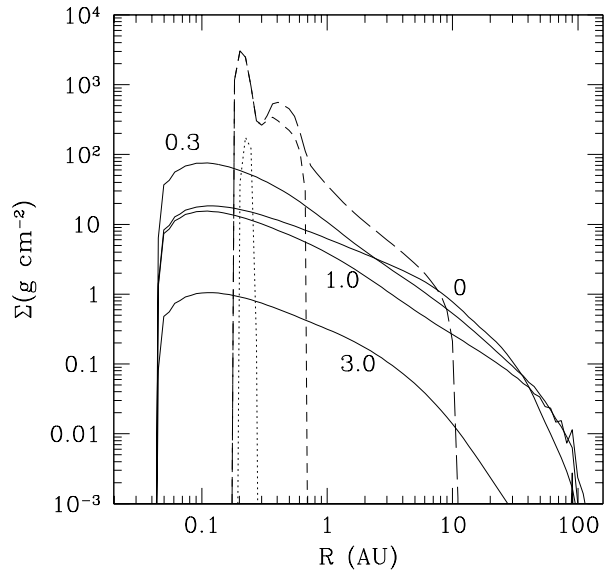}
\caption[X2.ps]{The solid curves indicate the dust density profile for our
composite dust model, at ages of 0, 0.3, 1 and 2~Myr. The dotted line is the
profile of formed planetesimals  after 0.3~Myr (the earliest deposition), the
short dashed line shows the planetesimals at 1~Myr and the long dashed
line shows the final deposition at 5~Myr. 
\label{X2}}
\end{figure}

Thus, we consider a model in which the solids 
are distributed in proportions of 10\% large (1~cm) particles, 20\% medium
(0.5~mm) particles and 70\% small (10 $\mu$m) particles. These proportions
are applied at all radii in the initial construction of the dust profiles. 
This is intended to model a more primitive dust distribution, to see how
the resulting planetesimal population is affected. The resulting distribution
is shown in Figure~\ref{X2}.


\begin{figure}
\includegraphics[width=84mm]{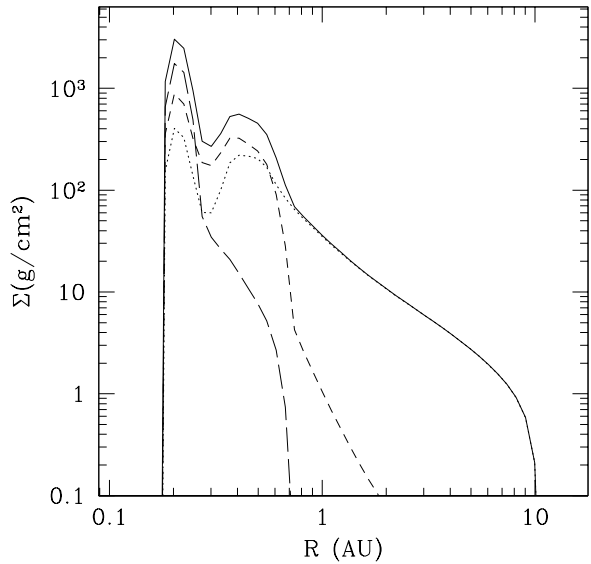}
\caption[Sizes.ps]{The solid curve indicates the surface density distribution
of material that was removed from our dust evolution due to gravitational instability.
The long dashed curve refers to the 1~cm particles, the short dashed curve to the
0.5~mm particles, and the dotted curve to the 10~$\mu m$ particles. We see that
solids that form on scales $> 1$AU are dominated by the smaller particles, while
the material at $\sim 0.1$AU is dominated by the larger particles.
\label{Size}}
\end{figure}

In  Figure~\ref{X2}, we see that the first planetesimals
to form are those at $\sim 0.2-0.3$~AU, after only 0.3~Myr. The composition of
these planetesimals is dominated by the large particles, whose profile is most
concentrated in the inner disc and whose radial mobility is responsible for
achieving the overdensity necessary for gravitational collapse. However,
by 1~Myr  we see that the planetesimal formation is starting to extend outwards
to the terrestrial planet domain, 
and eventually the
 final distribution extends out to $\sim 10$~AU.
Figure~\ref{Size} shows the distribution of sizes that are removed from the
dust disc by planetesimal formation over the full 5~Myr history of the disc.
We see that larger particles dominate the inventory at small scales, with
progressively smaller particles being incorporated further out and at later
times. 

In summary then, the choice of size distribution does not significantly
change the amount of mass distributed in planetesimals inside 1~AU, but
can influence the location of the first planetesimals to form (larger
particles move the location inwards) and how much mass is formed in
planetesimals are scales of several~AU (smaller particles enhance the
mass on larger scales).

\begin{figure}
\includegraphics[width=84mm]{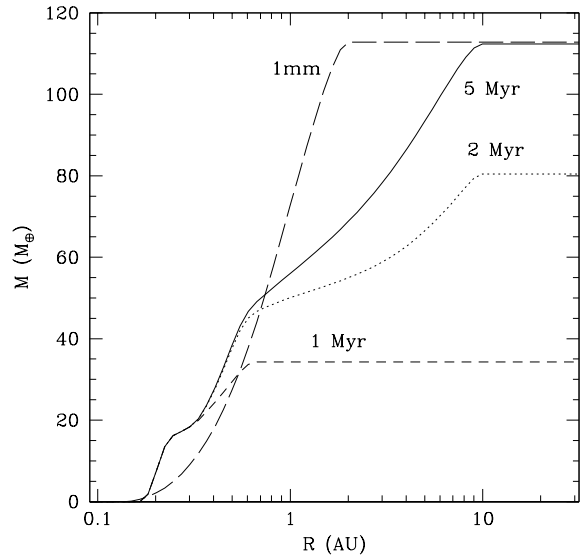}
\caption[Mass.ps]{The short dashed line indicates the mass deposited as planetesimals
interior to radius $R$, after 1~Myr of the discs evolution. The dotted line represents
the same quantity after 2~Myr, and the solid line represents the final deposition
profile, after 5~Myr, when the gas disc is largely depleted. The long dashed line
indicates the final planetesimal deposition in the case when the particles are all
of size 1~mm.
\label{Mass}}
\end{figure}

Figure~\ref{Mass} shows the enclosed mass in planetesimals as a function
of $R$ in the model of Figure~\ref{X2}. We see that $\sim 35 M_{\oplus}$ is deposited within
1~Myr inside 1~AU. After 2~Myr, this is supplemented by an additional
$45 M_{\oplus}$ that is deposited out as far as 10~AU. For comparison, we also
show the final mass deposition profile from the single size model in Figure~\ref{X1}. In
this case the bulk of the mass is deposited inside 2~AU.

 The mass inventory
in planetesimals laid down in this fashion may provide the starting conditions
necessary for a variety of planetary formation scenarios. The assembly simulations 
discussed in Hansen \& Murray (2012, 2013) require anywhere from 25--50 $M_{\oplus}$
of rocky material interior to 1~AU, depending on the mass inventory of the observed
system. Assembly of giant planet cores in the solar system requires 10--20 $M_{\oplus}$
of material on scales $\sim 3$--10~AU, which is also realised if there is sufficient
mass in particles smaller than 1mm.

\subsection{Wind Clearing}
\label{Wind}

We have also investigated an alternative model due to Armitage, Simon \& Martin (ASM), in which
the disc is cleared not by photoevaporation, but rather by the amplification of the
magnetic field in the disc and a resulting acceleration of the viscous evolution at late times.

\begin{figure}
\includegraphics[width=84mm]{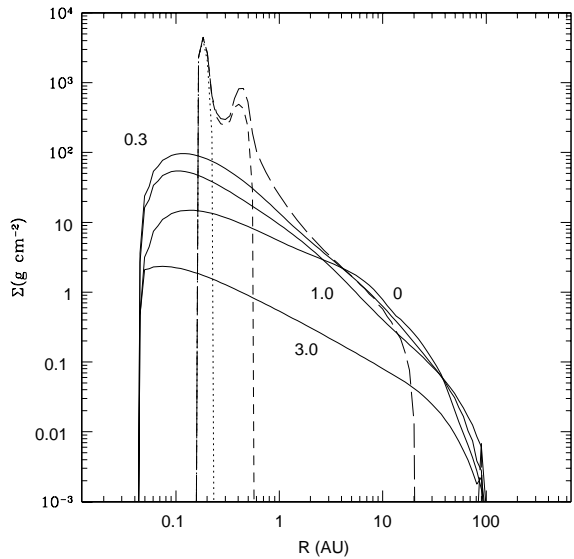}
\caption[Xbz.ps]{The solid curves indicate the dust profiles for
our composite size distribution at ages of 0, 0.3, 1 and 3~Myr. The
dotted line indicates the planetesimals formed within 0.3~Myr, the short
dashed line within 1~Myr and the long dashed line the final profile at 5~Myr.
The gas evolution in this case is calculated using the ASM model for disc
dispersion by magnetic winds, as described in the text.
\label{Xbz}}
\end{figure}

Our calculations suggest that the specific mechanism of disc clearing does not
qualitatively affect the mass deposition interior to 1~AU, but may influence
the amount of planetesimals formed on larger scales. Figure~\ref{Xbz} shows the
equivalent of Figure~\ref{X2}, but using the ASM model rather than the AA07 model,
with an initial parameterisation equivalent to the $\beta_z=10^5$ model of ASM (chosen
to clear the disc on a similar timescale to the photoevaporative model).
We include the same distribution of particle sizes, and find that the deposition
of planetesimals once again starts on scales of $\sim 0.2$AU after 0.3~Myr and
spreads slowly outwards on timescales of Myr. The only real difference between
the final planetesimal mass profiles is a slightly more extended disc at the
outer edge, driven by the fact that the late-time decrease in the gas surface
density is more rapid in this model.

\begin{figure}
\includegraphics[width=84mm]{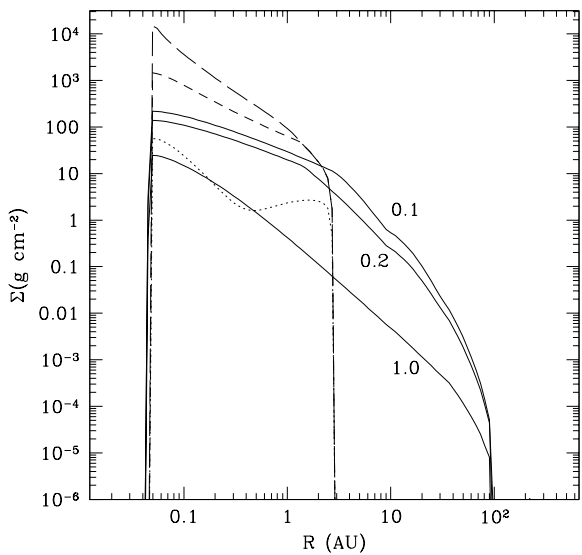}
\caption[Xbz.ps]{The solid curves indicate the dust profiles for
our composite size distribution at ages of 0.1, 0.2 and 1~Myr. The
dotted line indicates the planetesimals formed within 0.1~Myr, the short
dashed line within 0.2~Myr and the long dashed line the  profile at 1~Myr,
which is essentially the final model.
The gas evolution in this case is calculated using the ASM model for disc
dispersion by magnetic winds, while also removing the turbulent diffusion
contribution to the particle velocity.
\label{Xdiff}}
\end{figure}

The role of disc winds could have a larger effect on small scales if they
are capable of removing angular momentum without driving MHD turbulence.
Bai \& Stone (2013) find that turbulence can be suppressed if there is sufficient
vertical field threading the disk, with the disk wind able to carry away 
 the angular momentum necessary to facilitate accretion. Therefore, we have
run the same model as in Figure~\ref{X1} but with the diffusive contribution
to the particle velocity set to zero interior to 10~AU. The resulting evolution
is shown in Figure~\ref{Xdiff}. Two differences in the planetesimal deposition
are notable. The first is that the deposition occurs much more rapidly (starting
at 0.1~Myr and substantially finished by 1 Myr). The second is that the profile
extends inwards farther, indeed as far as the wind launching radius. 

Both features are the result of the fact that the turbulent diffusion acts
to smooth out a particle pileup -- acting against the inspiral exterior to
a local maximum in the dust density, but accelerating the inspiral interior
to it. Without this contribution, our dust evolution looks a lot more like
the model presented in Youdin \& Chiang (2004), wherein the required overdensity
for gravititional instability results purely from the concentration of the
solids towards the interior of the disk. However, the depletion of the
dust is again quite rapid and possibly at odds with observations of young
star forming regions.

\subsection{Imperfect Retention}
\label{Loss}

\begin{figure}
\includegraphics[width=84mm]{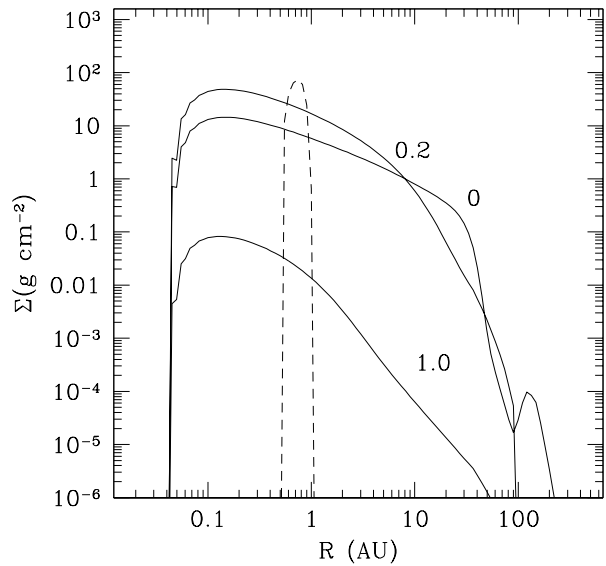}
\caption[X3.ps]{The solid curves again show the dust profile for our 0.1cm dust population
at 0, 0.2 and 1~Myr. In this case 50\% of the material is lost during each passage through the
inner disc. This has a dramatic consequence on the amount of material that forms planetesimals
via gravitational instability. The dashed line shows the profile such this material at 2~Myr.
This is much less than in Figure~\ref{X2} but would still be quite appropriate to the formation
of the solar system terrestrial planets.
\label{X3}}
\end{figure}

Our simple model assumes that the gas is slowly lost from the system, but that solids
circulate indefinitely. The result is that gravitational instability and planetesimal
formation is inevitable. How robust is this consequence in the face of the inevitably imperfect
retention of solids in a more realistic model? Some solid material may remain entrained with the gas that is
accreted onto the star, and some material may be carried to infinity by the outflow. 
Figure~\ref{X3} shows the evolution of the same model as Figure~\ref{X1} (dust contained
in 1~mm particles), but with a model which incorporates  mass loss.
In this case mass is lost from the inner edge of the disc
at the same rate as it is added back at the outer edge i.e. a 50\% retention fraction
with each cycle. Not surprisingly, the dust is depleted more rapidly, and with much
less planetesimal formation. However, the dashed line does show that some planetesimals do form,
with a concentration between 0.5--1~AU. The final amount of mass in solids is
$\sim 7 M_{\oplus}$, which actually makes an excellent initial condition for assembly of
systems like that of 
the terrestrial planets of our own solar system (Hansen 2009), although it lacks
the solid inventory on smaller scales that other systems require. Since the loss rate is
determined by the rate at which material passes through the inner edge of the disc, this
means that larger particles are lost more rapidly and smaller particles are retained
longer.

\begin{figure}
\includegraphics[width=84mm]{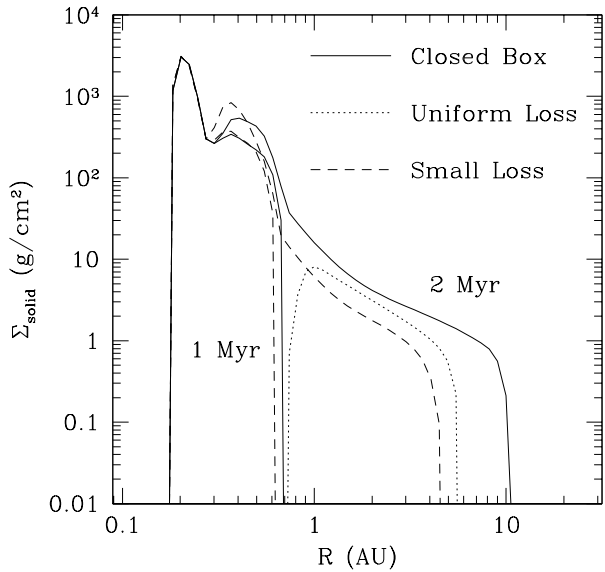}
\caption[Loss.ps]{The two solid curves indicate the material removed from the dust population
by gravitational instability in the closed box version of our model with a composite size population.
The curves are shown for system ages of 1~Myr and 2~Myr. The dotted curve is the equivalent 2 Myr for
a model in which 50\% of the material that passes through the inner disc is lost on each passage. There
is no dotted 1~Myr curve because no solids form that fast in this model.
The dashed curves shows the case when only the small particles (which still comprise 70\% by number)
are subject to the 50\% loss on each passage. Larger particles are retained, so that the results look
more similar to the closed box case.
\label{Losses}}
\end{figure}

Figure~\ref{Losses} shows the impact of dust loss on our simple model of a composite
population. The solid curve indicates the planetesimal deposition result from the closed box
model of Figure~\ref{X2}, at ages of 1~Myr and 2~Myr. The dotted line is the equivalent
for a model in which particles of all sizes are subject to the same 50\% loss on each passage
used in the model in Figure~\ref{X3}. We see that the biggest change is the reduction in the
amount of mass deposited into planetesimals at small radii. This is because these are composed
preferentially of the larger particles, which have shorter circulation times and therefore make
more passages through the inner disc per unit time. As such, they are preferentially depleted in
this simple model. Perhaps a more realistic model is to deplete only the smallest particles,
as Hu (2010) demonstrates that these are the most likely to be strongly coupled to the outflow
and lost from the system. The dashed line in Figure~\ref{Losses} indicates the consequence for
this model, in which the overall deposition is only slightly reduced from the closed box model.
This suggests that the broad features of the planetesimal deposition are robust to a moderate amount
of loss in the circulation and that the component of particles with sizes $\sim$ cm or larger are the
most likely to be depleted in the final accumulation. 

\subsection{Energy Limits on Circulation}

The recirculation of solids requires that gas drag from the outflow accelerate
particles until the outflow gas density drops low enough for the particle to
decouple from the drag force and fall back to the disk. Each time the gas lifts a particle out of the potential
well it saps the energy from the outflow. Although the total mass fraction of the
disk in solids is small, if dust particles make many passages through the disk
and are lifted each time by the outflow, then the energy required to lift the
dust to large radii many times may start to affect the energetics of the outflow.

For a single size particle distribution, with a characteristic disc traversal
time $\tau$, the particles will make $N \sim T/\tau$ passages through the disk
during a global lifetime T (which may be taken as 
the time to start forming planetesimals, which remove the dust from circulation). 
This means that gas drag from the wind must supply sufficient binding energy
to lift a mass $\sim N M_Z$ to the outer part of the disk. As long as this is
substantially less than the gas mass removed in the outflow, it will not have a
significant effect on the outflow dynamics. For our default model with mm-size
particles, $\tau \sim 
2 \times 10^4$~years and $T \sim 2 \times 10^5$~years,
 so that dust makes $\sim 10$ passages through the disk before being
absorbed into planetesimals. For global metallicities $\sim 0.01$, this means that
the outflow dynamics are not strongly affected as long as the outflow carries more
than 10\% of the gas mass that flows inwards from the disk to the star.

For dust populations with a size range, larger particles make more passages through
the disk in a finite time, and smaller particles fewer. Thus, for distributions
considered here (e.g. in Figure~\ref{X2}), with more of the mass in smaller particles, the
energy requirements for recirculation are well within the energy supplied by the
outflow. On timescales $> 0.1$Myr, the mass flux in the outflow must necessarily drop
as the gas disk mass drops, but the circulating dust mass also drops as the mass turns
into planetesimals on this timescale, so that the ability of the outflow to lift solid
material does not appreciably limit the end result of the circulation.

\section{Discussion}
\label{Observations}

The results of these calculations suggest that a protoplanetary disc with
a circulating dust component will naturally produce conditions
that allow for the formation of planetesimals via gravitational instability,
and with a substantial mass deposition interior to 1~AU. This offers
a promising route to the formation of planetary systems by in situ assembly
on these scales (Hansen \& Murray 2012; 2013; Chiang \& Laughlin 2013).
Of course, there are still several uncertainties that need to be addressed
in matching the properties of the initial planetesimal population to that of the 
final planetary system.

\subsection{Which Class of Planets are formed?}

\begin{figure}
\includegraphics[width=84mm]{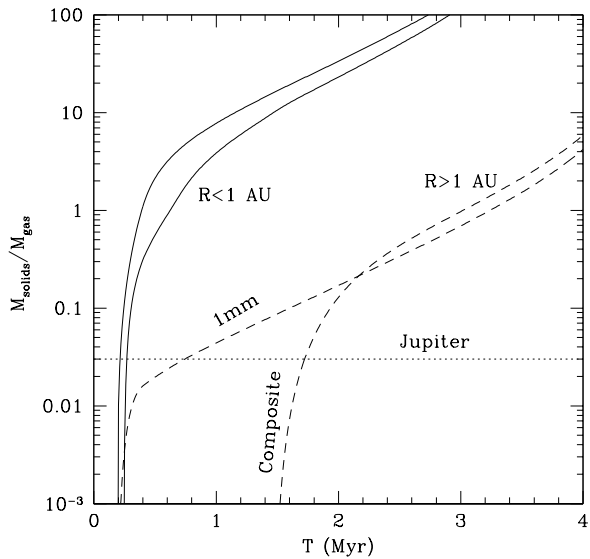}
\caption[InOut.ps]{The solid curve shows the ratio of total mass in planetesimals
to that of gas, interior to 1~AU as a function of time in our standard model. The
dashed line shows the equivalent quantity, but for the outer disc, defined as lying
 between 1~AU and 100~AU in this model. The horizontal dotted indicates the value for
Jupiter, assuming a total mass of 10$M_{\oplus}$ in heavy elements.
We see that formation of
gas giant planets requires formation within $\sim$ 1.8~Myr even with very efficient
accumulation of mass from the disc onto the planet. The gas mass in the inner disc
is exceeded by the planetesimal mass within 0.7~Myr, and reaches 1\% at 2.5--3 Myr.
\label{InOut}}
\end{figure}

The formation of planetesimals is a necessary step on the road to forming
a planet, but the conditions under which the subsequent assembly takes place
will determine what kind of planet ultimately results.
 The first planetesimals to form in this model do so
quite early, within 0.5~Myr of the original condensation from the nebula. This
occurs when there is still substantial gas present in the disc and thus the
future assembly into protoplanets will likely be affected by interactions with
the gas, which can affect both the mass inventory and potentially the radial
location. Figure~\ref{InOut} shows the relative mass in both gas and planetesimals
in both the composite size model and the one in which all dust is $\sim 1$~mm, for both the inner disc (defined here as $R< 1$AU) and the
outer disc ($1 AU<R<100 AU$). In the case of the inner disc, the amount of mass
in solids becomes comparable to the gas within 0.5~Myr.

 In models in which the disc clears via photoevaporation,
 such as those of AA07, the gas interior to 1~AU is not evaporated from the
system, but accreted onto the star. Thus, the eventual disposition of the remnant
gaseous disc will require it to either be accreted or cross the orbits of the 
final planets to get to the central object. As a consequence, this remnant gas
disc may have dynamical consequences for the final planetary configuration, but
it does not have sufficient mass to overwhelm the rocky inventory available for
forming planets. Indeed, much of this surviving gas may ultimately be accreted
by the planets rather than the star, since observations suggest that many of the
newly discovered planets are predominantly rocky but possess sufficiently
massive gaseous envelopes that they must have accreted some gas from the 
nebula (e.g. Lissauer et al. 2013). In the case of the model shown in Figure~\ref{X1},
the gas/planetesimals mass ratio is $\sim 3 \times 10^{-4}$ interior to 1~AU at
the point when photoevaporation decouples the outer gas disk from the inner gas
disk. Lopez \& Fortney (2013) suggest a characteristic value of the Hydrogen to rock
ratio is more like $\sim 1 \%$, which would require capture of the envelope on
slightly shorter timescales $\sim 3$~Myr (Figure~\ref{InOut}).

\begin{figure}
\includegraphics[width=84mm]{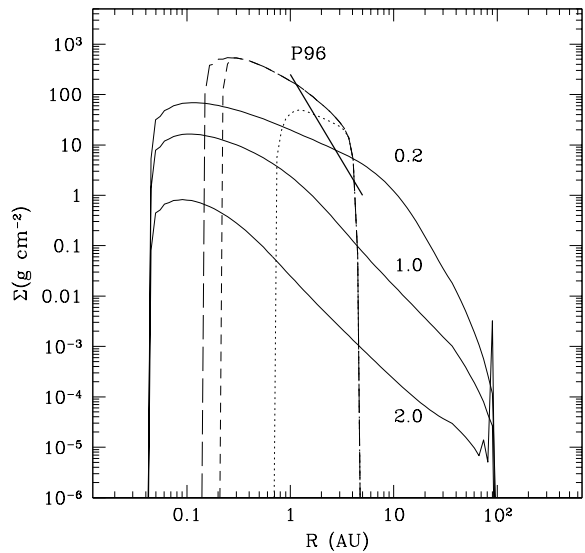}
\caption[XZ.ps]{The solid curves show the surface density in dust at ages
of 0.2, 1.0 and 2.0~Myr in a disk that starts with twice solar metallicity. 
The corresponding profiles of planetesimal deposition are indicated by the
dotted, short-dashed and long-dashed curves respectively. The thick solid
curve is the criterion identified by Pollack et al. (1996) as sufficient to
form the giant planet cores of our own solar system.
\label{XZ}}
\end{figure}

Exterior to 1~AU, there is a delay of 1--2~Myr before a population of planetesimals
begin to accumulate. Figure~\ref{InOut} demonstrates that the gas/rock ratio is more
variable on these scales, depending on the exact size distribution of the underlying
solids. Allowing for more massive discs can allow for the formation of rocky cores
more rapidly, but all models show the same requirement that the assembly of giant
planet cores needs to occur within a narrow window of 1--2~Myr, in order to capture 
sufficient gas from the nebula. This echoes the well-known constraint on the formation
of giant planets (e.g. Pollack et al. 1996). The process can be helped if we allow
for a higher metallicity in the gas. Figure~\ref{XZ} shows the same model as in
Figure~\ref{X1}, but we have increased the original dust sedimentation by a factor
of two to mimic an enhanced metallicity. We see that the planetesimal formation in
this model meets the surface density requirements of Pollack et al. (1996) within
0.2~Myr, suggesting that there is sufficient mass to form giant planet cores and
to accrete the gaseous envelopes. The fact that this process is enhanced in high metallicity
models may help to explain the correlation between host star metallicity and
giant planet frequency.

The last planetesimals to form are those on scales $\sim 10$~AU, which result from
the onset of gravitational instability during the last stages of the gas disc
clearing. The gas-poor origins of this outermost rocky component provides a natural
environment for the origin of Kuiper Belt analogues, and for the assembly of
the gas-poor 
ice giants of the outer solar system (e.g. Goldreich, Lithwick \& Sari 2004).
These effects may be further enhanced in cases where the evaporation is enhanced
by an external evaporation field (Throop \& Bally 2005).

\subsection{Aerodynamic Sorting}

The composition of the material that assembles into planetesimals in a composite model also
varies radially. Interior to $\sim 0.5$AU, the planetesimals are dominated
by larger particles ($\sim $cm sizes), while the planetesimals are dominated by medium sized
particles between 0.5--1~AU, and smaller particles exterior to that. This
natural aerodynamic sorting by particle size results from the different normalisations
of the steady state surface density distributions, as determined by the radial drift velocity.
Similar sorting is observed
in the chondritic material of the solar system (e.g. Scott \& Krot 2005), in which chondrite
families show an internal size consistency although there is some variation between
families, and has spurred models of sorting by turbulent processes (e.g. Cuzzi, Hogan \& Shariff 2008).
As we have shown, a dust population of this size can produce a planetesimal disc well
suited to the in situ assembly of compact planetary systems. However, composite size distributions
can also match this observation since mm-scale particles dominate the planetesimals that form
on scales $\sim$~1~AU, with larger(smaller) particles favouring planetesimals on smaller(larger) scales.

Furthermore, isotopic analysis of solar system materials also suggests that the process
of planetesimal assembly lasted for at least a few~Myr, in that chondrules are inferred
to be identifiably older (e.g. Connelly et al. 2008) than the oldest calcium-aluminum rich inclusions (CAI). This
is also nicely matched by the timescales in our model, with planetesimals at scales
$\sim $AU taking $> 1$Myr to form, although the innermost materials assemble within
0.5~Myr.

\begin{figure}
\includegraphics[width=84mm]{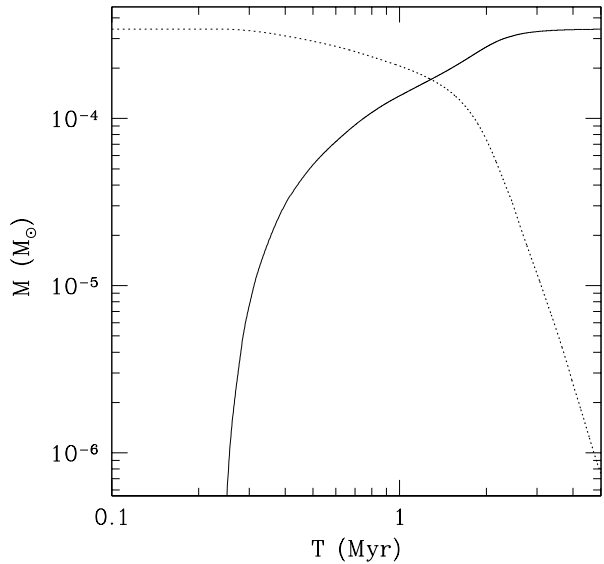}
\caption[Dust.ps]{The solid line indicates the amount of mass
contained in planetesimals in the model shown in Figure~\ref{X2}, while
the dotted line is the mass retained in the dust component. 
 We see that the dust mass remains substantial
for several Myr and only starts to decrease significantly once the
 outer parts of the disc
start to form planetesimals. 
\label{Dust}}
\end{figure}

\subsection{Evolution of Dust Disks}

Our baseline gas disc evolution is drawn from AA07, whose goal was to study the
evolution of dust that is observed in young protostellar systems through the
reprocessing of the stellar emission. This work was motivated by the observed
persistence of infrared emission from dust grains throughout the TT phase
(Andrews \& Williams 2005), contrary to the expectation that the mm-sized
dust grains should be removed by the radial drag forces discussed above
(Takeuchi \& Lin 2005). AA07 discuss replenishment by evolution and growth
in the grain population. While such processes are quite likely, our model 
also alleviates much of the observational discrepancy, since 
dust is not lost from the system but continuously recycled back to large radii. 
The formation of planetesimals will also remove dust (until such time as it
is replenished by planetesimal collisions), so that
 our model also
predicts a systematic loss of observable dust over time, shown in Figure~\ref{Dust}.
We see that the dust decreases by two orders of magnitude on a timescale $\sim 3$Myr, which is more palatable
than the timescale for loss via radial drift, which can be one to two orders of magnitude smaller. 

\subsection{Abundance Anomalies in Host Stars}

Another consequence of such a model is that the material accreted onto the star should be
depleted in heavy elements. Our nominal model contains an initial mass of 0.03$M_{\odot}$
of gas, most of which is ultimately accreted onto the star. For an original global metallicity
fraction $\sim 0.015$, this implies a total mass of $\sim 150 M_{\oplus}$ in potential
condensibles. However, using the condensation model of Hughes \& Armitage (2012), we only
condense 114 $M_{\oplus}$ into dust, with the difference being volatile material that
remains in the gas phase in the inner disc. If
we assume that only this remnant metal inventory is accreted with the gas, then we find that the material
accreted onto the star should slightly depress the observed overall metallicity and enhance the
volatile to refractories ratio slightly. To estimate
the amount of the  change, we assume that the surface convection zone during
this protostellar stage is $\sim 0.1 M_{\odot}$. If we add the above gas to this convection
zone, we decrease the net metallicity by $\sim 15\%$, or by 0.07~dex.

Some authors have claimed evidence for such metallicity trends (Ramirez et al 2009,
Ramirez et al. 2011, Gonzalez Hernandez et al. 2013, Ramirez et al. 2013) in nearby
stars, and have concluded that the Sun is depleted in refractories due to the formation
of the terrestrial planets. However, it is not clear how easy it is to define a 
suitable control set of stars without planets, given that observations suggest at least 20\%--50\% of sun-like stars
(Howard et al. 2010; Mayor et al. 2011) host low mass planetary systems, whose detectability
is not assured. Nevertheless, our results suggest that such enrichments are observable,
and do not require the assumption of late time accretion to avoid dilution of the signal,
because the material removed from the accreted gas is more than an order of magnitude larger
than that estimated using our, relatively low mass, terrestrial planet system (e.g. Chambers 2010).

\section{Conclusions}

We have presented a simple model for the evolution of a protoplanetary disc in which
dust particles undergo radial drift inwards, but are then recycled to the outer parts
of the nebula through the action of a stellar or disc wind. Although this model is quite
simplistic, it provides a natural framework for the deposition of tens of earth masses
of material into planetesimals on scales of 0.1--1~AU. This matches the required mass
inventory to assemble the observed planets in situ.

There are also a variety of subsidiary issues that suggest further study is warranted.
The retention of solids while gas is lost produces a natural evolution of the solid/gas ratio towards the limit where
gravitational instability and planetesimal formation is likely to set in, obviating
the need to invoke other physical mechanisms that require the existence of large
particles or anomalously low viscosities.
 Much of the planetesimal reservoir is deposited within 1~Myr, which allows for
the capture of residual gas from the nebula to explain the observed low mass Hydrogen
envelopes, and matches the timescales inferred from the cosmochemical age dating of
solar system meteoritic components. Furthermore. the gas mass on these scales is less
than the mass in planetesimals, so that the resulting planets are likely to be as
observed -- with substantial Hydrogen envelopes that are nevertheless a minority 
constituent by overall mass. Furthermore, we find that increasing the metallicity
of the disk has a larger effect on the mass of planetesimals formed on scales
of several AU, and thus provides a rationale for why the giant planet frequency
correlates with metallicity (Gonzalez 1997; Santos et al. 2004; Fischer \& Valenti 2005;
Johnson et al. 2010) more strongly than the frequency of lower mass planets
(Sousa et al. 2008; Bouchy et al. 2009; Mayor et al. 2011; Buchave et al. 2012). 

 If the solid retention is not perfect, and loss rate is size dependant, it provides an aerodynamic sorting mechanism
that may explain the characteristic sizes of chondrules in the solar system. Large
particles (dimensions of cm or larger) make more passages through the disc and are
thus likely to be more depleted via loss at the inner edge. Similarly, entrainment
in the outflows is more likely to remove small particles (Hu 2010), which suggests
that particles in the size range 0.01-1mm may have the greatest chances of retention
and survival.
 The circulation of solid material also helps to explain the apparent chemical homogeneity
of the solar system solid inventory (e.g. Villeneuve et al. 2009) and the ubiquity of
material processed at high temperatures (e.g. Brownlee et al. 2012).

There several ways in which this calculation could be improved. The size evolution of the dust component has been ignored,
although this is likely to provide an important feedback loop that may help to regulate the
radial profile of the eventual formed planetesimals. We have also not extended the model
forward to consider the formation of larger protoplanets and planets from our initial conditions.
Nevertheless, we consider the above results encouraging in the sense that they manage to
generate conditions that may plausibly be used to match to observed systems, at the reasonable price
of invoking an assumption that has already proven useful in other contexts and may be 
required anyway to explain the well-mixed compositions of solar system bodies.

The author thanks Phil Armitage, Andrew Youdin and the referee for helpful comments.
This research has made use of the NASA Exoplanet Archive, which is operated by the California Institute of Technology, under contract with the National Aeronautics and Space Administration under the Exoplanet Exploration Program.

\label{lastpage}
\newpage


\end{document}